# Label-dependent Feature Extraction in Social Networks for Node Classification

Tomasz Kajdanowicz[1], Przemysław Kazienko[1], Piotr Doskocz[1]

[1] Wrocław University of Technology, Wyb. Wyspiańskiego 27, 50-370 Wrocław, Poland
{tomasz.kajdanowicz, kazienko,piotr.doskocz}@pwr.wroc.pl

**Abstract.** A new method of feature extraction in the social network for within-network classification is proposed in the paper. The method provides new features calculated by combination of both: network structure information and class labels assigned to nodes. The influence of various features on classification performance has also been studied. The experiments on real-world data have shown that features created owing to the proposed method can lead to significant improvement of classification accuracy.

**Keywords:** feature extraction, label-dependent features, classification, social network analysis, AMD social network

## 1 Introduction

Classification is one of most important concepts in Machine Learning. It is usually based on the data that represents relationships between a fixed set of attributes and one target class. These relations describe each object independently that means no direct correlations between objects in the classification phase are taken into account. An exception may be additional input features, which aggregate information about the entire group a given object belongs to. However, it requires any clustering process be launched before. There are some applications and research methods, especially related to social networks, which are able to produce data with dependencies between labels of interconnected objects, referred as relational autocorrelation [16]. Based on these connections additional input information should be added to the classification process. If the considered objects are humans and the classification is utilized on their profiles then the social network can be extracted from complementary data (different from people's profiles) about common activities and mutual communication [9, 10, 15]. Overall, a social network is a set of nodes (human entities, objects) and node-node relationship between pairs of nodes [18]. According to [17], all network objects may be described by three distinct types of information that can be easily used in label classification: correlation between the object's label (class) and its attributes, correlation between the object's label and the observed (known) labels of other objects in its neighborhood and, consequently, correlation between the object's label and unobserved (unknown) labels of other objects in its neighborhood.

Basic task of within-network classification [1, 12] is to assign the correct labels to the unlabeled nodes from a set of the possible class labels. For example, based on the network of communication interactions, it could be determined whether a given company's employee is either an executive or a performer. Willing to obtain the best possible results of classification, all three types of information should be evaluated: nodes attributes (profiles), node-node network relations to the known labels in the neighborhood (labeled neighbors) and relations to the neighboring objects with unknown labels. Main difficulty here is to extract the set of most discriminative features from the network nodes and their connections to achieve the best classification model.

A new approach for network feature extraction is proposed in further sections. Some of these structural features have discriminative distribution, which may directly influence classification performance.

Section 2 covers related work while in Section 3 appears main part of the paper, where a new method for network feature extraction is presented. Sections 4 and 5, contain descriptions of the experimental setup and the obtained results, respectively. The paper is concluded in Section 6.

## 2 Related Work

In recent years, there has appeared a great number of works describing models and techniques for classification in network data. Analogously to classical machine learning problems, classification in network data requires specialized solutions for feature extraction, high performance supervised and unsupervised learning algorithms, sparse data handling, etc.

In general, network classification problems, may be solved using two main approaches: by within-network and across-network inference. Within-network classification, for which training entities are connected directly to entities, whose labels are to be classified, stays in contrast to across-network classification, where models learnt from one network are applied to another similar network [11]. Overall, the networked data have several unique characteristics that simultaneously complicate and provide leverage to learning and classification. More generally, network data allow the use of the features of the node's neighbors to label them, although it must be performed with care to avoid increase of variance estimation [7].

There have been developed many algorithms and models for classification in the network. Among others, statistical relational learning (SRL) techniques were introduced, including probabilistic relational models, relational Markov networks, and probabilistic entity-relationship models [2, 6, 13, 16]. Two distinct types of classification in networks may be distinguished: based on collection of local conditional classifiers and based on the classification stated as one global objective function. The most known implementations of the first approach are iterative classification (ICA) and Gibbs sampling algorithm (GS), whereas example of the latter are loopy belief propagation (LBP) and mean-field relaxation labeling (MF) [17]. Generally speaking, there exist many pretty effective algorithms of collective classification as well as graph-based semi-supervised learning methods. It refers, especially logForest, a logistic model based on links, wvRN, a relational neighbor model, SSL Gaussian random field model, ghostEdge, combination of statistical relational learning and semi-supervised learning for sparse networks and theirs collective classification supplements [5].

One of the most crucial problems in the network classification is feature extraction. According to [4] the derived features are divided into two categories: label-dependent (LD) and label-independent (LI). Features LD use both structure of the network as well as information about labels of the neighboring nodes labels, e.g. number of neighbors with given class label. Features LI, in turn, are calculated using the network structure only, e.g. betweenness of a node. The LI like features, therefore, are independent from the distribution of labels in the network and might not be informative. However, they can be perfectly calculated regardless of the availability of labels. What is worth mentioning, most of the proposed network classification methods were usually applied to the data sets with very limited access to labels. Their authors assumed that their applications need to deal even with only 1% labeled nodes. This problem is known as classification in sparsely labeled networks [4, 5].

It appears that the majority of network-based structural measures used as features in network classification may be useful and may potentially improve classification performance.

Social networks, being a network representation of interactions between people is a subject of research in terms of classification in networks as well [4].

## 3 Features Extraction from the Social Network

### 3.1 General Terms

Let us suppose that a social network is a graph $G = (V, E, X, L, Y, W)$, where $V$ is a set of nodes (objects, social entities); $E$ is a set of edges (connections) $e_{ij}$ between two nodes $v_i$ and $v_j$, $E=\{e_{ij}: v_i,v_j \in V, i \neq j\}$; $X$ is a set of attribute vectors $x_i$, a separate one for each node $v_i$ (a profile of $v_i$), $X=\{x_i: v_i \in V \Leftrightarrow x_i \in X\}$; $L$ is the set of distinct labels (classes) possible to be assigned to nodes; $Y$ is a list of actual labels assignments to nodes, $Y=\{<v_i,y_i>: v_i \in V \wedge y_i \in L\}$; $W$ is a set of edge weights, $\forall w_{ij} \in W\ w_{ij} \geq 0$ and $w_{ij}$ indicates the strength of edge $e_{ij}$.

Having known the values of $y_i$ for a given subset of nodes $V^K \subset V$, classification may be described as the process of inferring the values of $y_i$ for the remaining set of nodes $V^U$, $V^U = V \setminus V^K$.

The first step in the process of node classification is a translation of network data into a set of unified vectors, one for each node. A single vector corresponding to node $v_i$ contains all information from $x_i$ as well as some additional information (new attributes) derived by feature extraction methods based on the network profile. Next, the obtained set of vectors is used in classical, supervised classification.

### 3.2 Features Extraction

Feature extraction from social networks is a general term for methods of constructing variables from the connectivity graph, expressing the position and importance of each node with respect to the others. As mentioned in Section 1, the generated features may be label-independent or label-dependent. For clarity, while describing label-dependent features, it is made a basic assumption in the paper that feature extraction is based only on correlation between the object's label and the observed labels of other objects in its neighborhood see Fig. 1.

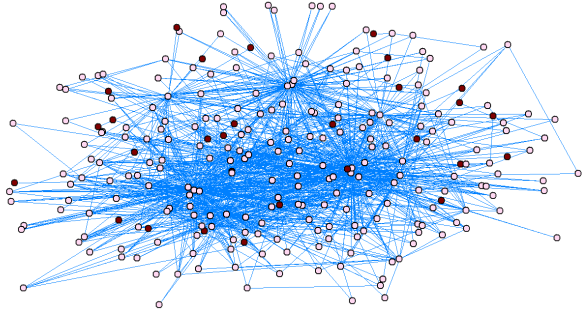

**Fig. 1.** Example social network with 10% of unlabeled nodes (black circles denote labeled nodes).

Three examples of basic label-independent and three label-dependent features are presented in the following sub-sections, as well as generalization for label-dependent features extraction.

#### 3.2.1 Label-independent Features

*Betweennes Centrality*
Betweenness centrality of node $v_i$ pinpoints to what extent $v_i$ is between other nodes. Nodes with high betweennes are very important in the network as other nodes are connected with each other mainly through them. Betweenness centrality $B(G,v_i)$ of node $v_i$ in graph $G$ can be calculated according to the following equation:

$$B(G,v_i) = \sum_{v_j,v_k,v_i \in G(V); j \neq k \neq i} \frac{P(G,v_j,v_k,v_i)}{P(G,v_j,v_k)}, \quad (1)$$

where:
$P(G,v_i,v_j)$ - a function returning the number of shortest paths between $v_i$ and $v_j$ in graph $G$;
$P(G,v_j,v_k,v_i)$ - a function that returns the number of shortest paths between $v_i$ and $v_j$ that pass through $v_i$ in graph $G$.
Obviously, Equation 1 is calculated only for pairs $v_j$, $v_k$, for which there exists a path from $v_j$ to $v_k$ to prevent the denominator from equaling 0.

*Degree Centrality*

Degree centrality is defined as the number of connections (edges) incident upon a given node. It is the simplest and most intuitive measures that can be used in the network analysis. Nodes with the high degree centrality are recognized as a crucial cog that occupies a central location in the network. Degree centrality $D(G,v_i)$ of node $v_i$ in graph $G$ can be computed using Equation 2:

$$D(G,v_i) = \frac{card(n(G,v_i))}{card(V)-1}, \quad (2)$$

where:
$n(G,v_i)$ - a set of neighboring nodes of node $v_i$ in graph $G$.

*Local Clustering Coefficient*

The local clustering coefficient $CC(G,v_i)$ of a node $v_i$ in graph $G$ quantifies how close $v_i$'s neighborhood is to a complete graph, see Equation 3.

$$CC(G,v_i) = \frac{card(R(n(G,v_i)))}{card(n(G,v_i))(card(n(G,v_i))-1)}, \quad (3)$$

where:
$R(V)$ - an operator returning the number of all connections between nodes from set $V$.

### 3.2.2 Label-dependent Features

While introducing label-dependent features two manners of their formation are proposed. Both of them relay on the idea of selective definition of sub-networks based on the labels assigned to each node. It means that a sub-network for a given label $l$ consists of only those nodes that share label (class) $l$ together with all edges connecting these selected nodes. For that purpose, a new selection operator $O(G,l)$ for graph $G$ and label $l$ is defined. It returns a sub-network $G_l$ labeled with $l$: $G_l=(V_l, E_l, X_l, \{l\}, Y_l, W_l)$ such that $V_l=\{v_i: <v_i,l>\in Y_l\}$, $Y_l=\{<v_i,y_l>: v_i\in V \wedge y_l=l\}$, $E_l=\{e_{ij}: v_i,v_j\in V_l \wedge e_{ij}\in E\}$, $X_l=\{x_l: v_l\in V_l \Leftrightarrow x_l\in X\}$.

Afterwards, for each sub-network $G_l$ (each label $l$), new features are computed.
First group of label-dependent features composition is based on new custom measures derived from the interaction between a given node and its neighboring nodes only. The measures take into consideration either the number of connections or their strengths.

*Normalized Number of Connections to the Labeled neighbors*

The measure for the normalized number of connections to the labeled neighbors $NCN(G,l,v_i)$ represents the proportion of the number of connections to the neighboring nodes in the sub-network with label $l$ ($G_l$) by the number of connections to the labeled neighbors in the whole primary graph $G$ (with all labels).
The measure $NCN(G,l,v_i)$ is defined as follows:

$$NCN(G,l,v_i) = \frac{card(n(O(G,l),v_i))}{card(n_L(G,v_i))}, \quad (4)$$

where:
$n(O(G,l),v_i)$ - a set of the neighboring nodes for node $v_i$ in sub-network $O(G,l)$,
$n_L(G,v_i)$ - a set of $v_i$'s labeled neighbors in graph $G$, each neighbor must be labeled with any label $l \in L$.

Note that the value of $card(n(O(G,l),v_i))$ is the same as the number of connections between $v_i$ and $v_i$'s neighbors (each $v_i$'s neighbor has one connection with $v_i$). Similarly, the value of $card(n_L(G,v_i))$ equals the number of connections between $v_i$ and all $v_i$'s labeled (and only labeled) neighbors.

The measure $NCN(G,l,v_i)$ is computed separately for each label $l$ and in general, for two labels $l_k$ and $l_m$, the value of $NCN(G,l_k,v_i)$ may differ from $NCN(G,l_m,v_i)$.
For the example network from Fig. 2, and the measure $NCN(G,'red',v_1)$ calculated for node 1 in the sub-network with nodes labeled with the *'red'* class, the value of $NCN(G,'red',v_1)$ is 4 divided by 8 (total number of nodes in graph $G$).

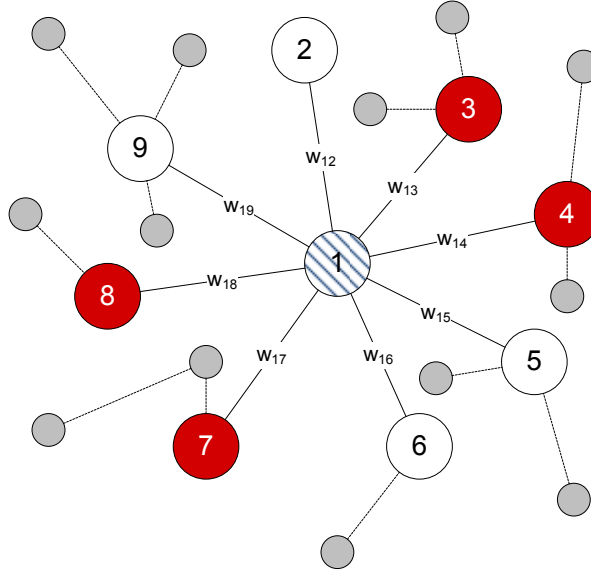

**Fig. 2.** Feature calculation based on label dependent neighborhood. For each of label class {white, red} $w$ is calculated.

*Normalized Sum of Connection Strengths to the Labeled Neighbors*
The value of the normalized sum of connection strengths to the labeled neighbors $NCS(G,l,v_i)$ is the proportion of node $v_i$'s activity towards $v_i$'s neighbors (measured by the aggregated connection strengths) in the sub-network with label $l$ ($G_l$) normalized by the equivalent value of strengths to the neighbors with any label in the whole graph $G$. The value of $NCS(G,l,v_i)$ for graph $G$ and label $l$ is expressed in the following way:

$$NCS(G,l,v_i) = \frac{\sum_{v_j \in n(O(G,l),v_i)} w_{ij}}{\sum_{v_j \in n_L(G,v_i)} w_{ij}}, \tag{5}$$

Similarly to $NCN(G,l,v_i)$, the measure $NCS(G,l,v_i)$ is evaluated separately for each label $l$ and differs for different labels $l$.

For the network from Fig. 2, the measure $NCS(G,'red',v_1)$ is computed for node 1 and label (class) *'red'*, as the sum of $w_{13}$, $w_{14}$, $w_{17}$, and $w_{18}$ normalized by sum of all eight connection strengths.

### 3.2.3 General Method for Label-dependent Features Extraction

In the domain of social network analysis (SNA), a number of measures characterizing network nodes have been introduced in the literature. Majority of them is label-independent and it is possible to define many methods that will extract label-dependent features based on them. A general concept of creation of any label-dependent feature $M_l(G,l,v_i)$ for label $l$ and node $v_i$ in the social network $G$ applies label-independent feature $M$ to the appropriate labeled sub-network $G_l=O(G,l)$, as follows:

$$M_l(G,l,v_i) = M(G_l,v_i), \tag{6}$$

where:
$M_l(G_l,v_i)$ - denotes any structural network measure for node $v_i$ applied to sub-network $G_l=O(G,l)$, e.g degree, betweennes or clustering coefficient;

Obviously, $M_l(G,l,v_i)$ is computed separately for each label $l$ using the appropriate sub-network $G_l=O(G,l)$. In other words, Eq. 6 provides a method for construction of label-dependent version of certain metric.

As an example, the label-dependent clustering coefficient ($CC_l$) is defined in accordance with Equation 3 as:

$$CC_l(v_i) = \frac{card(R(n(G_l,v_i)))}{card(n(G_l,v_i))(card(n(G_l,v_i))-1)} \tag{7}$$

## 4 Experimental Setup

### 4.1 Data Set

The data set used for experiments, "Attendee Meta-Data" (AMD), was downloaded from UCI Network Data Repository (http://networkdata.ics.uci.edu/data.php?d=amdhope). The AMD data set was an output of a project, which used RFID (Radio Frequency Identification) technology to help connect conference participants at "The Last HOPE" Conference held in July 18-20, 2008, New York City, USA. All attendees received an RFID badges that uniquely identified and tracked them across the conference space. The data set contains descriptions of interests of participants, their interactions via instant messages, as well as their location over the course of the conference. Conference attendees were asked to "tag" themselves based on a diverse set of interests. Thanks to location tracking, a list of attendances was extracted for each conference talk. Additionally, participants could email or send a text message to "ping" the people who had similar interests.

In general, the data set contains information about conference participants, conference talks and presence on talks. Initial import contained 767 different persons, 99 talks, 10,110 presences reported during talks. In the cleaning process, these contributors who did not give any information about their interests were excluded from further studies. As a result, 334 persons with 99 lectures and 3,141 presences have left after cleaning.

Afterwards, the social network was build. Ties in the network were constructed based on the fact that participants were present on the same talks. Moreover, strengths of the connections between each pair of contributors were calculated as the proportion of number of talks attended by both participants by the total number of talk presences of the first participant. It provided 68,770 directed, weighted connections, with histogram presented in the Fig. 3.

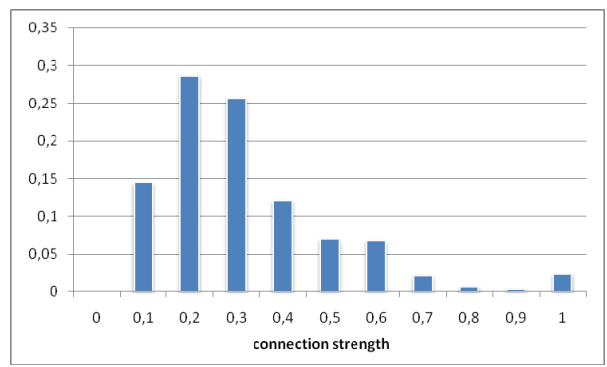

**Fig. 3.** Histogram of calculated weights in the AMD social network.

The raw data contained 4 attributes: 3 nominal (sex, cell phone provider, country) and 1 numerical (age). Additionally, each participant was described by unordered set of interests that in our experiments was chosen as the classification target. Since each network node (participant) could have multiple interests assigned, it was decided to construct 20 separate experimental data sets that formed a binary assignment of each interest. Example networks are presented in Fig. 4 and 5. For the clarity of the experiment, the binary classification problem was established as it did not contrive a loss of generality of the proposed feature extraction approach.

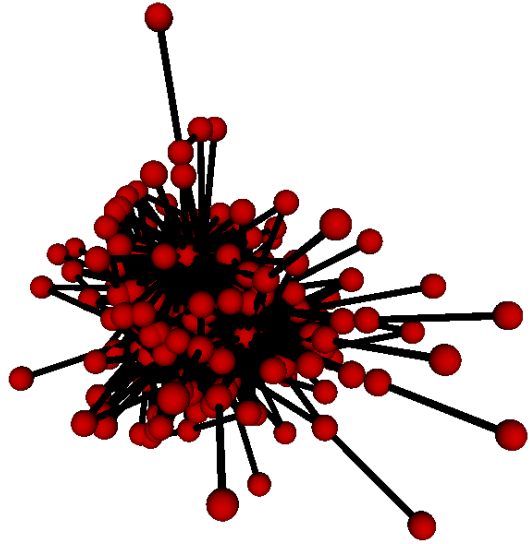

**Fig. 4.** Visualization of the social network for the activism interest data set based on the class '0' neighborhood using Force-Directed Placement Algorithm [3].

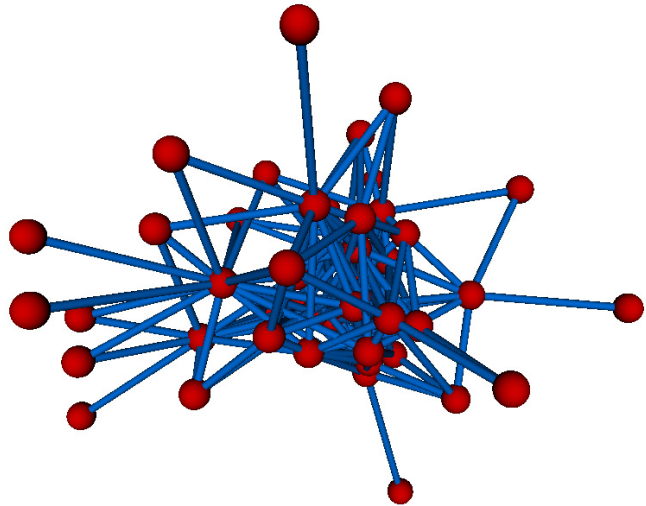

**Fig. 5.** Visualization of the social network for the activism interest data set based on the class '1' neighborhood using Force-Directed Placement Algorithm [3].

### 4.2 Extracted features

According to the methodology presented in Section 3, 17 attributes were calculated in the experiments, see Table 1.

**Table 1.** Features used in experiments.

| No. | Feature | Feature Set |
|---|---|---|
| 1 | age | |
| 2 | gender | 1 |
| 3 | county | |
| 4 | phone provider | |
| 5 | betweenness, Eq. 1 | |
| 6 | degree, Eq. 2 | 2 |
| 7 | clustering coefficient, Eq. 3 | |
| 8 | normalized sum of connection strengths to the neighbors labeled as '0', Eq. 5 | |
| 9 | normalized sum of connection strengths to the neighbors labeled as '1' neighbors, Eq. 5 | |
| 10 | normalized number of connections to the neighbors labeled as '0' neighbors, Eq. 4 | |
| 11 | normalized number of connections to the neighbors labeled as '1', Eq. 4 | |
| 12 | betweenness on neighborhood with class '0' | 3 |
| 13 | betweenness on neighborhood with class '1' | |
| 14 | degree on neighborhood with class '0' | |
| 15 | degree on neighborhood with class '1' | |
| 16 | clustering coefficient on neighborhood with class '0' | |
| 17 | clustering coefficient on neighborhood with class '1' | |
| | all above (1-17) | 4 |

Extracted features were grouped in 4 sets. The first contained raw data attributes. In the second there were label-independent network based features. In the third group label-dependent features obtained from proposed method were introduced. The last, fourth group attach all previously introduced features. Finally, the obtained 20 data sets, used in the experiment, may be downloaded from http://www.zsi.pwr.wroc.pl/~kazienko/datasets/amd/amd.zip in the arff format.

The outcome of performed classification (classification target) was established to predict an interest that a particular person has assigned.

### 4.3 Classification

Experiments were conducted for 20 data sets using 3 classification algorithms, AdaBoost, Multilayered Perceptron, SVM, with settings presented in Table 2, the same for each of four feature groups (Table 1). Classification was performed in 10% - 90% proportion of labeled and unlabeled nodes, respectively, using 10-cross fold validation.

**Table 2.** Features used in experiments.

| Algorithm | Setting | Value |
| --- | --- | --- |
| AdaBoostM1 | weight threshold | 100 |
|  | number of iterations | 10 |
|  | base classifier | Decision Stump |
| Multilayer Perceptron | learning rate | 0.3 |
|  | momentum | 0.2 |
|  | training time | 500 |
|  | validation threshold | 20 |
|  | hidden layers | 5 |
| SVM | complexity | 1.0 |
|  | tolerance | 0.0010 |
|  | epsilon | $10^{-12}$ |
|  | kernel | polynomial kernel |
|  | exponent | 1.0 |

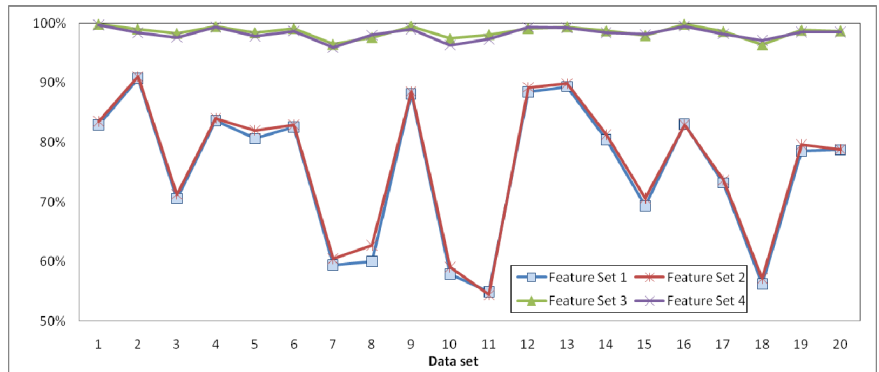

**Fig. 6.** Average accuracy for 20 data sets using 4 different feature sets.

## 5   Results

The obtained results have revealed that the average accuracy of classification using various feature sets really differs. As presented in Fig. 6, the average accuracy is greater by about 23% for feature set 3 and 4 compared to set 1 and 2. Simultaneously, F-Measure and precision improves by usage of label-dependent feature sets (set 3 and 4) by 33% and 35%, respectively, see Table 3.

Irrespectively of the used feature data set, all utilized classification algorithms: AdaBoost, Multilayered Perceptron, SVM, provide similar results (see Fig. 7).

As shown in Fig. 6, classification based on feature set 3 and 4 seems to be more stable than for feature set 1 and 2. In particular, standard deviation of accuracy for 20 data sets in first case equals 1% and in the second 12%.

Additionally, experiments have revealed that classification based on feature set 4 returns in average worse accuracy than classification based on feature set 3 (see Table 3). Let remind that feature set 4 contains all features from sets 1, 2 and 3. Worse classification performance might be an effect of too many relative poor input features, from which some weaken classification and have contrary discriminative distributions. It refers features from set 1 and 2 that degrade high correlation between output and label-dependent features from set 3. It means that the features extracted from the social network are so good that regular profiles of the tested cases only decrease classification performance and should not be even taken into account.

Owing to the carried out experiments, it is visible that the proposed label-dependent features used in classification undoubtedly provide the best results.

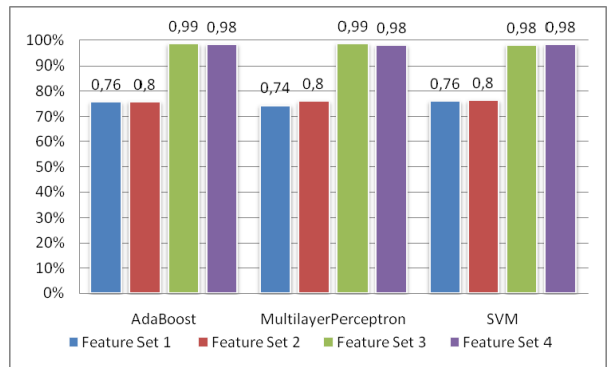

**Fig. 7.** Average classification accuracy for 4 different feature sets and 20 data sets.

**Table 3.** Average results of experiments for 20 data sets.

| Algorithm | Feature Set | | | | Measure |
|---|---|---|---|---|---|
| | 1 | 2 | 3 | 4 | |
| AdaBoost | 0.76 | 0.76 | 0.99 | 0.98 | Accuracy |
| | 0.62 | 0.63 | 0.99 | 0.99 | Precision |
| | 0.67 | 0.68 | 0.99 | 0.98 | F-measure |
| Multilayer Perceptron | 0.74 | 0.76 | 0.99 | 0.98 | Accuracy |
| | 0.67 | 0.63 | 0.99 | 0.98 | Precision |
| | 0.69 | 0.68 | 0.99 | 0.98 | F-measure |
| SVM | 0.76 | 0.76 | 0.98 | 0.98 | Accuracy |
| | 0.64 | 0.61 | 0.98 | 0.98 | Precision |
| | 0.69 | 0.67 | 0.98 | 0.98 | F-measure |

## 6 Conclusions and Future Work

A new method for label-dependent feature extraction from the social network was proposed in the paper. The main principle behind the method is based the selective definitions of sub-graphs for which new features are defined and computed. These new features provide additional quantitative information about the network context of the case being classified.

According to collected experimental evidences, the proposed label-dependent feature extraction appears to be significantly more effective and improves classification performance in high extent. Obtained, so good, results were even surprising to authors. These results have shown that the new approach to classification extended with features derived from the social network may return very satisfactory and promising outcomes.

It may even happen that the regular features only decrease classification indicators and should be removed from the input feature set. This phenomenon comes probably from the general background of both feature sources. Human profiles are, in fact, the voluntarily collected data whereas social networks are created upon real people activities. There is a crucial difference between a statement "I am interested in mountains" and real information about the mountain climbing. The second is more reliable.

Feature work will focus on further experimentations on the method, especially in terms of its validity for variety of local network measures. Additionally, the proposed feature extraction method will also be examined against the usage of global objective functions for classification. Yet another direction of future studies will be development of new ensemble algorithms, which would have network measures already incorporated, especially based on boosting concept [8].

**Acknowledgments.** This work was supported by The Polish Ministry of Science and Higher Education, the development project, 2009-11